%% file: article.tex
\documentclass[reprint,prl,superscriptaddress]{revtex4-1}
\usepackage{booktabs}
\usepackage{amsmath}
\usepackage{graphicx}
\usepackage{color}
\usepackage{tikz}
\usepackage{longtable}
\def\Heff{H_\mathrm{eff}}
\def\Jeff{J_\mathrm{eff}}
\def\Jpeff{J'_\mathrm{eff}}

\begin{document}
 
\title{Long-range doublon transfer in a dimer chain induced by topology 
and ac fields}
 
\author{M.~Bello}
\email{miguelbellogamboa@gmail.com}
\affiliation{Instituto de Ciencias de Materiales, CSIC,
Cantoblanco, E-28049, Madrid, Spain} 
\author{C.E.~Creffield}
\affiliation{Departamento de F\'isica de Materiales, Universidad
Complutense de Madrid, E-28040, Madrid, Spain}
\author{G.~Platero}
\affiliation{Instituto de Ciencias de Materiales, CSIC,
Cantoblanco, E-28049, Madrid, Spain} 

\date{\today}


\begin{abstract}
The controlled transfer of particles from one site of a
spatial lattice to another is essential for many tasks in
quantum information processing and quantum communication. 
In this work we study how to induce long-range
transfer between the two ends of a dimer chain,
by coupling states that are localized just on the chain's end-points.
This has the appealing feature that the transfer occurs only
between the end-points -- the particle does not pass through the
intermediate sites -- making the transfer less susceptible to decoherence.
We first show how a repulsively bound-pair of fermions, known as a doublon,
can be transferred from one end of the chain to the other
via topological edge states.
We then show how non-topological surface states of the familiar Shockley
or Tamm type can be used to produce a similar form of transfer under
the action of a periodic driving potential. Finally we show that
combining these effects can produce transfer by means of more exotic
topological effects, in which the driving field can be used to 
switch the topological character of the edge states, as measured by
the Zak phase.
Our results demonstrate how to induce long range transfer of strongly 
correlated particles by tuning both topology and driving.
\end{abstract}

\maketitle

Recent experimental advances have provided reliable and tunable setups to 
test and explore the quantum mechanical world. Paradigmatic examples are 
ultracold atomic gases trapped in optical lattices and coherent semiconductor 
devices such as quantum dots. Much of the interest in the last few years has 
been focused on the long-range transfer of particles in these systems, bearing 
in mind potential applications in the fields of quantum information and 
quantum computing. Several mechanisms have been proposed to achieve this
aim, including propagation along spin chains 
\cite{bose} or a bipartite lattice \cite{bipartite},
coherent transport by adiabatic passage (CTAP) 
\cite{greentreeChip, greentreeCharge, greentree,busch}, 
or the virtual occupation 
of intermediate states \cite{busl, Braakman, sanchez}.
Harnessing the effects of topology has also recently become possible,
in which edge states provide lossless transport that
is protected against disorder. Key to the production of these topological
insulators has been the use of time-dependent potentials to engineer
the tunnelings in these lattice systems. This has allowed the production
of quantum Hall states \cite{aidelsburger} \cite{miyake}, and more exotic topological
systems such as the Haldane model \cite{haldane}.
It has also been shown that driving graphene with ac electric fields
can be used to induce a  semimetal insulator transition \cite{alvarografeno}.
Inspired by these developments, in this work we study how the long-range
transfer of particles can be achieved by combining these ingredients;
topological effects and periodic driving.

Probably the most simple system that can exhibit topological effects
is the one-dimensional dimer chain, or one-dimensional Su-Schrieffer-Heeger
(SSH) model, originally introduced to describe solitonic effects in
polymers \cite{ssh} \cite{zhang}. Such a dimer chain supports edge states 
when it is in the topologically non-trivial phase. This is 
determined by the ratio between the two hopping rates, $J$ and $J'$, 
a parameter we will call $\lambda=J'/J$ \cite{ssh,delplace,alvaro}. Recently, 
many investigations have focused on this model and several results have been 
confirmed experimentally using ultracold atoms trapped in optical lattices 
\cite{nat}. Since these edge states form a non-local two-level system,
a remarkable dynamics can occur for non-interacting particles moving on 
such a chain; they can directly pass from one end to the other without 
moving through the intermediate sites \cite{supp}. 
This direct transfer of 
particles between distant sites, which preserves the quantum coherence
of the state, clearly has applications to quantum information
processing, in which quantum states must be coherently shuttled between
quantum gates and registers. 

In this work we investigate how this long-range transfer of particles 
in a dimer chain can 
be produced and optimized in systems of strongly interacting fermions.  
In general, interactions are known to destroy the topological effects 
in the non-interacting dimer chain \cite{supp}. However, by
considering the strongly-interacting limit, in which fermions form
repulsively-bound pairs called ``doublons'' \cite{winkler,charles_chain,charlesRombi}, 
we show that the effect can be recovered by tuning local potentials at the end-points
of the lattice. We further show that driving the system with a high-frequency
potential allows the manipulation of the doublon tunneling rates via 
the phenomenon known as coherent destruction of tunneling \cite{hanggi},
permitting
an alternative form of long-range transfer to occur via a non-topological 
mechanism that we term ``Shockley transfer''. Finally we show how combining 
lattice topology with the driving potential gives rise to transfer
via exotic topological effects, giving extremely fine control over
process of long-range doublon transfer. 

\begin{figure*}[t]
\centering
\includegraphics[width=\textwidth]{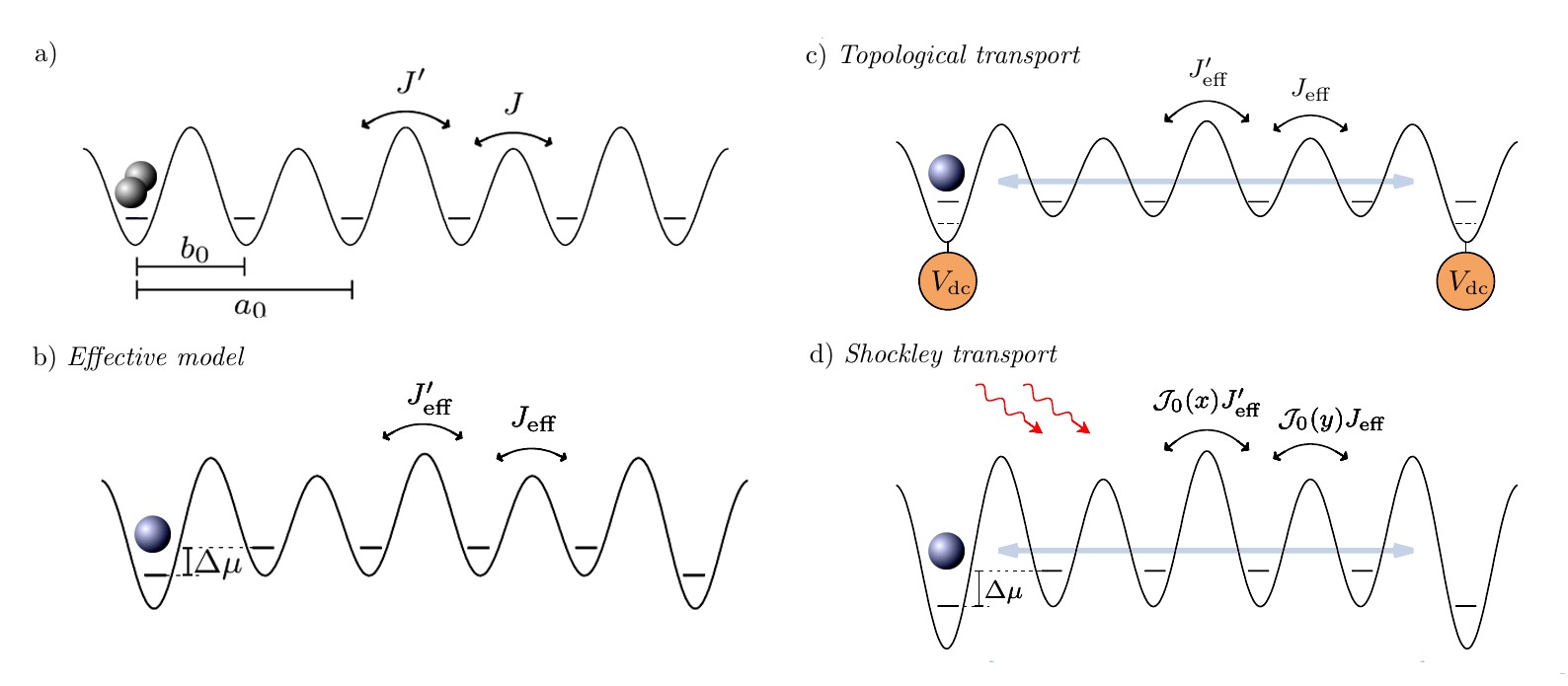}
\caption{{\bf Schematic representations}
a) The full Hamiltonian (\ref{full_ham}).
A chain of M dimers characterized by two hopping rates, $J$ and $J'$, 
the lattice constant, $a_0$, and the intra-dimer distance, $b_0$. 
The other important parameter of the model is the interaction strength, 
$U$, which needs to be large enough with respect to the hoppings for 
doublons to form. 
b) Scheme displaying the main features of the effective model 
Hamiltonian (\ref{eff_mod}). We consider the doublon as a single 
quasiparticle which moves through the lattice with hoppings 
$\Jpeff=2J'^2/U$ and $\Jeff=2J^2/U$. In a finite system, 
a chemical potential difference arises between the endpoints and the rest 
of the lattice sites. 
c) Scheme showing how to produce topological long-range transfer of doublons. 
A gate potential at the terminating sites of the chain is needed to restore 
the lattice periodicity. 
d) Scheme showing the Shockley long-range transfer of doublons. 
The ac-field renormalizes the doublon hoppings to $\mathcal{J}_0(y)\Jeff$ 
and $\mathcal{J}_0(x)\Jpeff$, where $x=\frac{2Eb_0}{\omega}$ and 
$y=\frac{2E(a_0-b_0)}{\omega}$, leaving the chemical potential unaffected.}
\label{scheme}
\end{figure*}

\section{Results}

\subsection{Doublon dynamics} In
the limit of strong interactions, fermions on a lattice can pair to form stable
bound states known as ``doublons'', even if the interaction is repulsive.
This effect is a consequence of the discretization of space; the kinetic
energy of a particle is limited by the width of the energy band, and so if
the interaction energy is sufficiently large, the decay of the doublon into
free particles is
forbidden on energetic grounds. Doublons have been observed in several
systems such as ultracold atomic gases \cite{winkler} 
and in organic salts \cite{wall}.
 
The system we have studied can be modelled by a SSH-Hubbard Hamiltonian:
 \begin{multline}
  H = -J' \sum\limits_{i=1, \sigma}^M 
c_{2i \ \sigma}^\dagger c_{2i-1 \ \sigma} - 
  J \sum\limits_{i=1, \sigma}^{M-1} 
c_{2i+1 \ \sigma}^\dagger c_{2i \ \sigma} + H.c.\\
   + U \sum\limits_{i=1}^{2M} n_{i \uparrow} n_{i \downarrow}
  = H_{J}+H_{U} \quad ,
\label{full_ham}
 \end{multline}
where $c_{i \ \sigma}^\dagger$ ($c_{i \ \sigma}$) is the standard
creation (annihilation) operator for a fermion of spin
$\sigma$ on site $i$, and $n_{i \ \sigma} = c_{i \ \sigma}^\dagger c_{i \ \sigma}$
is the number operator. The hopping Hamiltonian $H_J$ is parameterized
by the two hopping parameters $J$ and $J'$  which describe the dimer
structure of the lattice (shown schematically in Fig. \ref{scheme}a),
while $H_U$ accounts for the interactions between particles
by a Hubbard-$U$ term.
 
\begin{figure}
\centering
\input{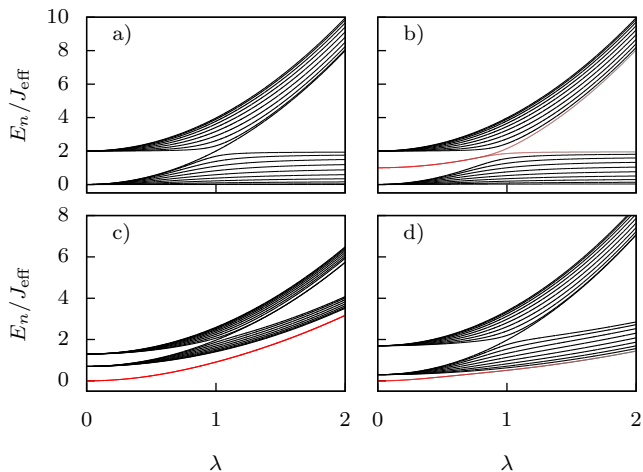}
\caption{{\bf Energy levels of a 10 dimer chain}, 
$b_0 = a_0/2$, the energy is measured in units of $\Jeff$. 
a) Without a gate potential, $\Delta\mu=-2 J^2/U$, there are no states outside 
the bulk bands, and therefore no edge states for any value of $\lambda$. The 
interaction destroys the edge states as long as the system is in the 
strongly-interacting regime and no ac field is applied.
b) Adding a gate potential to compensate for $\Delta\mu$, so that 
$\Delta\mu + \mu_{\mathrm gate}=0$. The two states with energies in the gap 
(red lines) for $\lambda<\lambda_c$ are the edge states predicted by the topology 
of the system. For $\lambda \geq \lambda_c$ the gap closes, and the system 
becomes topologically trivial.
c) Driven by an ac-field with intensity and frequency such that 
 $\mathcal{J}_0 (E a_0/\omega)=0.3$, two localized states separate from 
the bottom of the lowest band (red lines). We can see how the ac field 
renormalizes the hoppings, making the bands narrower and increasing their 
separation from the Shockley states.
d) Same case with $\mathcal{J}_0 (E a_0/\omega)=0.7$.}
\label{spectrum}
\end{figure}

We study the two fermion case, the smallest number of fermions
that can form a doublon, and restrict ourselves to the
singlet subspace (one up-spin and one down-spin).
In order to obtain an effective Hamiltonian that accurately models the 
dynamics of doublons, we can perform a unitary transformation perturbatively 
in powers of $J/U$ and $J'/U$ \cite{supp,hofmann} (see Methods). 
Assuming we only have one doublon 
in the system, we can neglect interaction terms between doublons and the 
hopping processes of single particles, to arrive at 
 \begin{multline}
  \Heff= \Jpeff \sum\limits_{i=1}^M d_{2i}^{\dagger} d_{2i-1} 
  + \Jeff \sum\limits_{i=1}^{M-1} d_{2i+1}^{\dagger} d_{2i} + H.c. \\
  + \sum\limits_{i=1}^{2M} \mu_i n_{i}^d \quad ,
\label{eff_mod}
 \end{multline}
 \begin{equation}
  \mu_i =  \left\{ \begin{array}{lcc}
                 \mu_\mathrm{bulk}=\Jpeff+\Jeff+U & \text{if} & 1<i<2M\\
                 \mu_\mathrm{edge}=\Jpeff+U & \text{if} & i\in \{1,2M\}
                   \end{array} \right.
 \end{equation}
Here $\Jeff=2J^2/U$ and $\Jpeff=2J'^2/U$. 
$d_i^\dagger=c_{i\uparrow}^\dagger c_{i\downarrow}^\dagger$, ($d_i$) is 
the creation (annihilation) operator for a doublon on site $i$, and 
$n_i^d=d_i^\dagger d_i$ is the doublon number operator. In the effective model 
(\ref{eff_mod}), the doublon hopping rates are smaller than the original ones 
and positive regardless of the sign of $J$ or $J'$ (see Fig. \ref{scheme}b). 
Unexpectedly, this transformation also gives rise to
a chemical potential term, $\mu_i$, which depends on 
the number of neighbors of site $i$. While all sites in the bulk of the chain
have two neighbors, the two end-sites have only one and thus
experience a different value of $\mu_i$, which breaks the lattice periodicity.
The difference in chemical potential is given by
$\Delta\mu=\mu_\mathrm{edge}-\mu_\mathrm{bulk}=-\frac{2J^2}{U}$.
In Fig. \ref{spectrum}a we show the energy spectrum for a chain of 10 dimers,
where we can clearly see that even when the system is topologically
non-trivial ($\lambda < \lambda_c$, see below), 
no edge states are visible. This is a consequence
of this finite-size effect; the alteration in chemical potential at the
ends of the lattice causes the edge states' energies to enter the bulk bands.
As a consequence the system does not support edge states for doublons.
This corroborates the result that interactions destroy topological transfer.

\subsection{Topological transfer}
Analogously to the non-interacting case, the topology in the present case is determined by 
the ratio between the effective hoppings, given by 
$\Jpeff/\Jeff=(J'/J)^2=\lambda^2$. 
For an infinite chain, the system is in the topologically non-trivial phase
when $\lambda  < \lambda_c = 1$; for the finite case of $M$ dimers the 
critical value of the ratio is given by 
$\lambda_c=\sqrt{1-\frac{1}{M+1}}$ \cite{delplace}. 

To obtain topological transfer for doublons, we
must restore the lattice periodicity by adding a gate voltage, 
$\mu_\mathrm{gate}$, to the edge sites to compensate for the
difference in chemical potential, such that 
$\Delta\mu + \mu_\mathrm{gate} = 0$. In this way we recover 
edge states for a chain with doublons. 
We show the result in Fig. \ref{spectrum}b, and we can indeed see that
the two edge states lie between the bulk bands for $\lambda <\lambda_c$.

\begin{figure}
\centering
\input{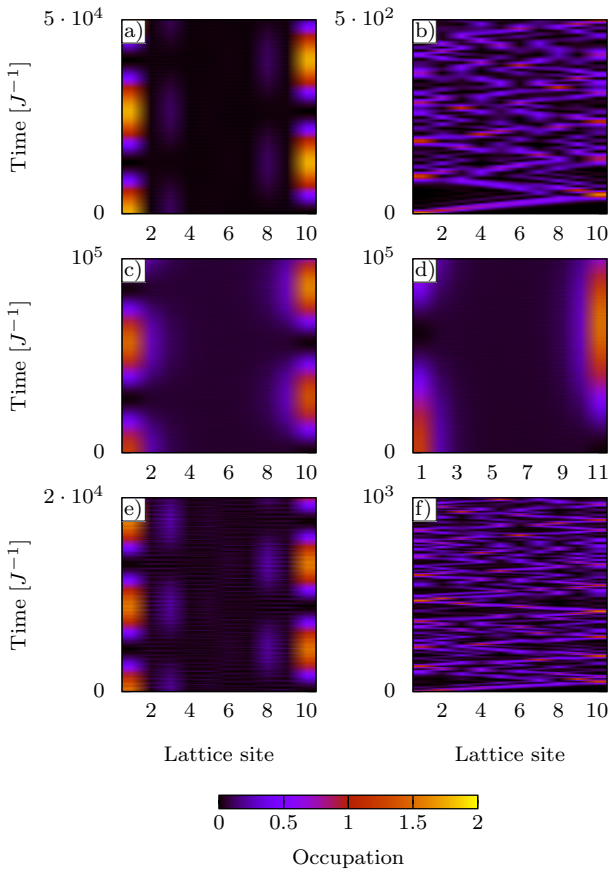}
\caption{{\bf Time evolution of the site occupation.}
In all cases $U=16J$ and the initial condition consists of two fermions in a singlet state 
occupying the first site of the chain. The simulations are for a chain containing 5 
dimers except for (d). 
\textit{Topological transfer}: 
a) Chain with compensating gate potentials at the edge-sites and $\lambda=0.5<\lambda_c$ 
(topological regime). The doublon oscillates from one edge to
the other without occupying intermediate sites, giving an example of long-range
topological transfer in an interacting system. 
b) As before, but with $\lambda=1>\lambda_c$ (trivial regime). The doublon 
now simply spreads over the entire lattice. 
\textit{Shockley transfer}: 
c) Chain driven by an ac field with parameters $\lambda=1$, $b_0=a_0/2$, 
$E/\omega=1.6/a_0$ and $\omega=2J$. By using the ac field to renormalise
the effective hoppings, we can obtain long-range transfer without
compensating the chemical potentials of the edge points.
d) AC driven chain with same parameters, but an odd number of sites. Long-range transfer is 
mediated by the Shockley mechanism (no topological transfer would be possible in this case).
\textit{AC induced topological transfer}: 
e) AC driven system with compensating gate potentials, 
parameters are $\lambda=1.2$, $b_0=0.6a_0$, $2E/\omega=3.6/a_0$ and 
$\omega=2J$ (topological regime, red square in Fig. \ref{zak_plot}b). Long range
transfer occurs unlike  in the undriven system ($\lambda=1.2$). f) As in (e) but with $2E/\omega=2 a_0^{-1}$ 
(trivial regime, green dot in Fig. \ref{zak_plot}b). As expected, no long 
range transfer occurs.}
\label{dynamics}
 \end{figure}

We show examples of the dynamics in Figs. \ref{dynamics}a and \ref{dynamics}b;
in the topological regime the doublon oscillates between the two edge-sites
without passing through intermediate sites, whereas in the trivial regime the
doublon simply spreads over the entire lattice.
Interestingly, due to the sublattice symmetry of the system, 
when the number of sites is odd, there is one and only one edge state in the 
chain, localized on one end or the other depending on whether 
$\lambda<\lambda_c$ or $\lambda>\lambda_c$ \cite{supp}. 
Thus, there is 
{\em no} long-range doublon transfer for systems with an {\em odd} number of 
sites (half-integer number of dimers).

\subsection {Shockley transfer}
The effective Hamiltonian for doublons (\ref{eff_mod}) contains, as we 
discussed above, a site-dependent chemical potential which breaks 
translational symmetry. This produces Shockley-like surface states 
\cite{shockley} if the hopping rates $\Jeff$ and $\Jpeff$ 
are smaller than $|\Delta\mu|$. Usually this is not the case, however 
there is an efficient way to induce such states by driving the system 
with a high-frequency ac-field. The ac-field renormalizes the hoppings \cite{hanggi} 
which become smaller than in the undriven case. This cannot be achieved,
for example, by simply reducing the hoppings $J$ and $J'$ by hand, since this 
will also affect the effective chemical potential
which still will be of the same order of $\Jeff$ and $\Jpeff$. 
To model the driven system we add a periodically oscillating potential that 
rises linearly along the lattice
\begin{equation}
\begin{split}
  H(t) & =H_J+H_U+E\cos \omega t \sum\limits_{i=1}^{2M} x_i (n_{i,\uparrow}
  +n_{i,\downarrow})\\  & \simeq \Heff
  +E\cos \omega t \sum\limits_{i=1}^{2M} 
  x_i (n_{i,\uparrow}+n_{i,\downarrow}) \ ,
 \end{split}
\label{time_H}
 \end{equation}
where $E$ and $\omega$ are the amplitude and frequency of the driving,
and $x_i$ is the spatial coordinate along the chain. Since the Hamiltonian 
(\ref{time_H}) is periodic in time, $H(t)=H(t+T)$, we can apply Floquet 
theory and seek solutions of the Schr\"odinger equation of the form 
$\lvert\psi(t)\rangle=e^{-i\epsilon_n t}\lvert\phi_n (t)\rangle$, where $\epsilon_n$ 
are the so called Floquet quasienergies, and $\lvert\phi_n (t)\rangle$ are a 
set of $T$-periodic functions termed Floquet states. Quasienergies play 
the same role in the time evolution of the system as conventional
energies do for a static Hamiltonian. In the strongly interacting regime, a 
perturbative calculation shows that the hopping terms
are renormalized by the zeroth Bessel function (see Methods and \cite{supp}),
$\Jeff\rightarrow \mathcal{J}_0 (y)\Jeff$, and 
$\Jpeff\rightarrow \mathcal{J}_0 (x) \Jpeff$, 
where $y=\frac{2E(a_0-b_0)}{\omega}$ and $x=\frac{2Eb_0}{\omega}$  

\cite{charles_chain, alvaro}. We show the effect of this renormalization
in Fig. \ref{spectrum}c; as the effective tunneling
reduces in magnitude the bulk bands becomes narrower, and the Shockley
states are pulled further out of them. The factor of $2$ in the 
argument of $\mathcal{J}_0$ comes from the doublon's twofold electric 
charge. The geometry, which so far has not played any role, now becomes 
important in this renormalization of the hoppings. The simplest case
is for $b_0=a_0/2$, in which both hoppings are renormalized by the same factor 
$\mathcal{J}_0(Ea_0/\omega)$. 
An important point is that the on-site effective 
chemical potential, being a local operator, commutes with the periodic 
driving potential, and so is not renormalized. This is the critical reason
for using a periodic driving to modify the tunneling; it renormalizes
the values of $\Jeff$ while keeping the chemical potential unchanged. 
 
\begin{figure}
\centering
\input{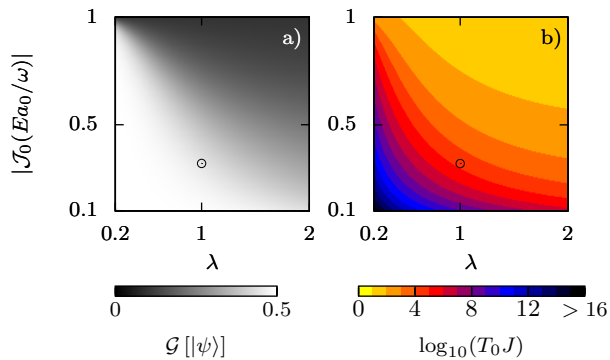}
\caption{{\bf Characterizing Shockley transfer.}
a) Correlation, $\mathcal G$, between the edge occupancy of the 
Shockley-like surface states in an ac-driven chain containing 5 dimers. 
We have considered the case $b_0=a_0/2$. The long-range transfer occurs 
in the pale region (lower-left) of the parameter space. 
b) Transfer time, $T_0$, computed as $\pi/\Delta\epsilon$, 
where $\Delta\epsilon$ is the energy difference between the two edge states. 
$T_0$ tends to infinity as $\mathcal{J}_0(Ea_0/\omega)$ or $\lambda$ go to zero.
The black circles correspond to the parameters of the time evolution shown 
in Fig. \ref{dynamics}c; the transfer time, $T_0\sim 10^4 J^{-1}$, is correctly 
reproduced. As can be seen, a slight change in the 
field parameters can change the transfer time by several orders of magnitude.}
\label{correlation}
 \end{figure}

We show in Figs. \ref{spectrum}c,\ref{spectrum}d how 
varying the hopping rates has the effect 
of pulling two energies out of the bulk bands, inducing the presence of 
localized states at the edges. These edge states occur in pairs and 
so also form a non-local two level system \cite{shockley}. Nevertheless
they can be affected by local perturbations and so unlike the previous
case, are topologically unprotected \cite{beenakker}. 
From the stationary eigenstates of Hamiltonian (\ref{eff_mod}) with 
renormalized hoppings, we can define a quantity, 
$\mathcal{G}\left[\lvert\psi\rangle\right]$, 
that measures the density correlation between the end-sites 
for a given eigenstate, $\lvert\psi\rangle$,
\begin{equation}
 \mathcal{G}\left[\lvert\psi\rangle\right]:=| \langle 1 | \psi \rangle
\langle N | \psi \rangle |,\quad
  \mathcal{G}\in[0,1/2] \ .
 \end{equation}
Here $\lvert i\rangle=d^\dagger_i\lvert 0\rangle$ is the basis of localized doublon states. 
If the total occupancy at the ends, 
$| \langle 1 | \psi \rangle |^2 + | \langle N | \psi \rangle |^2$,
 is a constant then $\mathcal{G}$ is maximum when
$| \langle 1 | \psi \rangle | = | \langle N | \psi \rangle |$.
In addition, the energy difference between the two edge states tells us
how fast the doublon transfer time is, $T_0 = \pi / \Delta \epsilon$. 
We can see in Fig. \ref{correlation}a that when the values of the hoppings 
are reduced by the ac field, the two lowest energy eigenstates of 
Hamiltonian (\ref{eff_mod}) become more localized at the edges. 
Smaller values of $\lambda$ favor localization as well. This produces cleaner 
dynamics with less unwanted occupancy of the intermediate sites of the chain. 
On the other hand, we can see in Fig. \ref{correlation}b that the transfer 
time rapidly increases, soon becoming too large to observe
in simulations or in experiment.
At larger values of the hoppings the edge states enter the bulk bands, as can be seen in  
Fig.\ref{spectrum}d close to $\lambda=1$, and the 
long-range transfer of doublons is suppressed. 
 
We show examples of the dynamics for a periodically-driven system in 
Figs. \ref{dynamics}c, \ref{dynamics}d. 
Since the origin of the edge states is not topological, long-range transfer 
can occur via this mechanism
for chains  even with an odd number of sites, as seen in Fig. \ref{dynamics}d.

\subsection{AC induced topological transfer}
If we combine both methods, adding a gate potential at the ends {\em and} 
driving the system with an ac-field, it is possible to bring the system 
into exotic topological phases. The effective Hamiltonian is
simply given by \eqref{model_2} without the chemical potential term.

There is a close connection between the correlation of the edge-occupancy,
$\mathcal G$, for those states which close the gap,
and the Zak phase, $\mathcal{Z}$ \cite{zak}. This is the topological invariant that 
classifies 1D Hamiltonians with time reversal, particle-hole and chiral 
symmetry. The Zak phase has already been calculated for a driven dimer 
chain without interaction \cite{alvaro}, and it is straightforward to 
extend it to our effective model for doublons 
 \begin{equation}
  \mathcal{Z}=\frac{\pi}{2} \left[ 1+
\mathrm{sgn} \left( \mathcal{J}^2_0(y)-
\lambda^4\mathcal{J}^2_0(x) \right) \right]  \ .
\label{order_parameter}
 \end{equation}

\begin{figure}[!htb]
\centering
\input{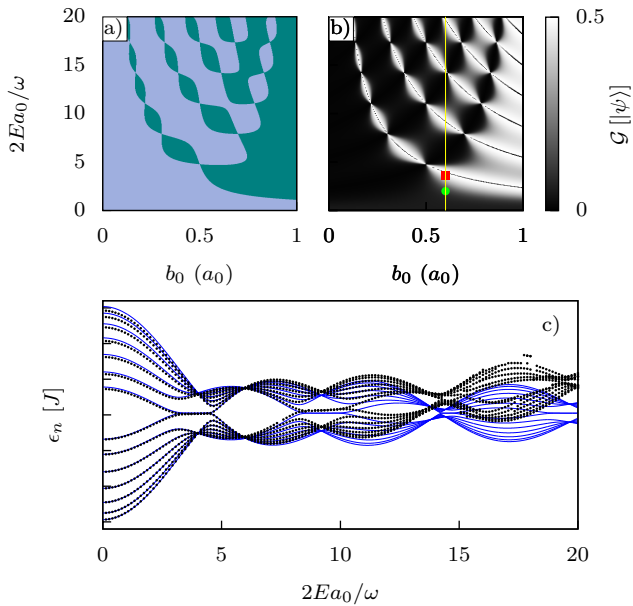}
\caption{{\bf Exotic topological transfer.}
a) Plot of $\mathcal{Z}$ (Eq. \ref{order_parameter}) for $\lambda=1.2$; 
green regions are those in which the system is in the  
topologically non-trivial phase ($\mathcal{Z}=\pi$). 
b) Plot of $\mathcal{G}\left[\lvert \psi \rangle \right]$ 
computed for a chain containing 7 dimers with the same $\lambda$ as in a). 
The black lines that cross the white regions correspond to the zeros of 
$\mathcal{J}_0(2Eb_0/\omega)$. At those points in  
parameter space, the different sites of the chain are uncoupled; the system is 
thus degenerate and  $\mathcal{G}$ is not a well-defined quantity. 
The yellow line marks the cross-section through parameters space
used in the quasienergy plot.
The red square and green dot  
mark the parameters of the system for
the time evolutions shown in Figs. \ref{dynamics}e and f respectively. 
c) Quasienergies as a function of the driving 
amplitude $E$. The other parameters are set to $U=16J$, $b_0=0.6 a_0$ 
and $\omega=2J$. The edge states appear as predicted by the phase diagram, 
and correspond to the highest values of $\mathcal{G}$. For large values of 
the ac-field intensity, the doublon states begin to couple with the other 
states of the system and the exact quasienergies ( black dots) 
diverge from those predicted by the effective model (blue lines).}
\label{zak_plot}
 \end{figure}

The argument of the Bessel functions is twice that for a non-interacting 
system, and the factor $\lambda^4$ comes from the squared ratio between 
the effective doublon hoppings. In Figs. \ref{zak_plot}a, \ref{zak_plot}b 
we compare the
phase diagram obtained by plotting (\ref{order_parameter}) and the result 
obtained by computing $\mathcal{G}\left[\lvert\psi\rangle\right]$. We can see that the
agreement is excellent, indicating that the Zak phase can be directly
measured from the density correlation function. In Fig. \ref{zak_plot}c
we show the quasienergy spectrum for $b_0 = 0.6 a_0$, to make a cross-section
through the parameter space. It can clearly be seen that when the system 
is topologically non-trivial, corresponding to ${\mathcal G} \simeq 0.5$,
a pair of edge states emerges from the bulk bands and enters the gap. When the
system is topologically trivial they then reenter the bulk again.
The dots in Fig. \ref{zak_plot}c 
were obtained from the 
diagonalization of the unitary time-evolution operator for one period of 
the full original Hamiltonian \eqref{time_H} with an added gate potential, 
making no approximations.
The agreement between these quasienergies, and those calculated
from the effective model (\ref{model_2}) is extremely good for
driving parameters $2 E a_0 / \omega \leq 10$, indicating that
our approximation schemes are valid. For larger values of the driving
parameters our effective model still captures the behaviour of
the quasienergies, but small deviations begin to appear as the doublon states
begin to couple with other states of the system.

In Figs. \ref{dynamics}e  and \ref{dynamics}f we show two examples of the
dynamics corresponding to the two points marked
in Fig. \ref{zak_plot}b. When the system is topologically non-trivial
(red square) the system exhibits long-range doublon transfer as expected.
In the topologically trivial regime (green dot), however, this does not
occur, and the doublon instead propagates throughout the whole lattice.

 \section{Discussion}

We have derived an effective Hamiltonian for two particles in a 
quantum dimer chain that 
bind together via a repulsive interaction. Interestingly this binding
produces an effective surface potential, different from that of the bulk.
In general this surface potential prevents topological transfer of
particles, but by adding local gate potentials
to compensate for it, topological transfer can be recovered.
We have also shown that by adding a periodic driving potential to
renormalize the hoppings, while leaving the surface potential unchanged,
we can produce long-range transfer via Shockley states. This
transfer, is, however, not topologically protected.
Finally, by combining topological transfer with an ac driving
field, we can obtain a rich topological phase diagram, in which long-range
transfer occurs when the Zak phase is non-zero.

It is natural to ask how this long-range transfer phenomena depend 
on the total number of dimers forming the chain. The transfer time is 
essentially the inverse of the energy difference in the two-level 
system formed by the hybridization of the edge states. This energy 
difference is related to the overlap between the edge states, 
which are solutions that decay exponentially from the surface of the lattice. 
Thus we can conclude that the transfer time increases exponentially when 
increasing the size of the chain. Another fact which affects the transfer 
time is the dependence of $\Jpeff$ and $\Jeff$ on $U$. Increasing the 
interaction strength has the effect of slowing down the dynamics.

Ultracold atoms confined in optical lattice potential are extremely
clean and only slightly affected by decoherence. In units of the tunneling time, 
doublon life time in a three dimensional optical lattice, has been found to 
depend exponentially on the ratio of the on-site interaction to the kinetic energy \cite{demler}.
This is not in general 
the case for electron transfer in semiconductor nanostructures where
hyperfine or spin-orbit orbit interactions induce decoherence, which 
strongly depends on the material.  However, since we deal with doublons,
forming a singlet state, 
spin relaxation and decoherence is suppressed by the energy difference 
between the intradot singlet and excited triplet states. 

In summary, we propose three ways for long-range transfer of 
strongly-interacting particles, all mediated by edge states. In the first 
case, non-trivial topological edge states are required. In the second, 
long-range transport is mediated by Shockley states induced by ac driving. 
Finally, combining both topology and driving allows us to tune the range of 
parameters where  long-range transfer is achieved.
Our proposal could be experimentally confirmed both in cold atoms and in 
semiconductor quantum dot arrays. In these last systems either charge 
detection by means of a quantum detector, such as a quantum point contact 
or an additional quantum dot, or transport measurements are within 
experimental reach.

Our results open  new  avenues to achieve direct transfer of interacting 
particles between distant sites, an important issue for quantum information 
architectures.

\section{Methods}

\subsection{Effective Hamiltonian for doublons} 
The energies of a one-dimensional lattice form a Bloch band with
a width of $2J$, and thus the maximum kinetic energy carried
by two free particles is $4J$. If the particles are initially
prepared in a state with a potential energy much greater
than $4J$, the initial state then cannot decay without the
mediation of dissipative processes. We consider the regime where for 
doublons to split is energetically unfavorable, i.e. \! $U \gg J,J'$. 
Following \cite{hofmann} we obtain an effective Hamiltonian just for the 
doublons in a dimer chain by means of a Schrieffer-Wolff (SW) transformation, 
projecting out the single-occupancy states. This transformation is performed 
perturbatively in powers of $J/U$ and $J'/U$ and up to second order gives 
rise to the effective Hamiltonian \eqref{eff_mod} where the hoppings 
$J$ and $J'$ become renormalized by the interaction. 

\subsection{AC driven Hamiltonian: hopping renormalization}
If one wants to deal with interactions between particles as well as 
interactions with an external driving, and treat both on an equal footing, 
a more elaborate procedure than before is necessary. 
For a time-periodic Hamiltonian $H(t+T)=H(t)$ with $T=2\pi/\omega$, 
the Floquet theorem states that the time evolution operator $U(t_2,t_1)$ 
can be written as:
\begin{equation}
 U(t_2,t_1)=e^{-iK_\mathrm{eff}(t_2)}e^{-i\Heff(t_2-t_1)}e^{iK_\mathrm{eff}(t_1)} \,
\end{equation}

with a time-independent effective Hamiltonian, $\Heff$ governing 
the slow dynamics and a $T$-periodic operator $K_\mathrm{eff}(t)$ that accounts 
for the fast dynamics; $\exp(-iK_\mathrm{eff}(t))$ is also termed the 
\textit{micromotion operator}. In the high-frequency limit, by which we 
mean $\omega\gg \Jpeff$ and $\Jeff$, these operators can be expanded in powers 
of $1/\omega$:
\begin{equation}
 \Heff=\sum\limits_{n=0}^\infty \Heff^{[n]}\text{ ,}\quad 
 \Heff^{[n]}\propto \bigg(\frac{1}{\omega}\bigg)^n
\end{equation}
For a detailed description of the high-frequency expansion (HFE) 
method see \cite{universal,andre,mikami}. 
Now we express our periodic Hamiltonian \eqref{time_H} in the rotating frame with 
respect to both the interaction and the ac field \cite{bukov_guay}:
\begin{align}
& H_\mathrm{int}(t) =\mathcal{U}^\dagger(t)H(t)
\mathcal{U}(t)-i\mathcal{U}^\dagger(t)\partial_t\mathcal{U}(t) \ , \nonumber\\
& \mathcal{U}(t)=\exp\big(-iH_U t-i\textstyle\int H_{AC}(t) dt\big) \ .
\end{align}
Here $\int H_{AC}(t) dt$ is just the antiderivative of the operator: 
\begin{equation}
 H_{AC}(t)=E\cos \omega t \sum\limits_{i=1}^{2M} x_i (n_{i,\uparrow}
  +n_{i,\downarrow}) \ .
\end{equation}
To derive the effective Hamiltonian, we perform the HFE of $H_\mathrm{int}(t)$ up 
to first order in $1/\omega$, see \cite{supp}. It can be seen that in the limit 
$U\gg\omega$ the result is the same as the one obtained by first performing the SW 
transformation and then the hoppings renormalization. 
Conversely, in the limit $\omega\gg U$ the result is consistent 
with first doing the high-frequency hopping renormalization and then the SW 
transformation. This coincidence can be understood since the great difference 
between $U$ and $\omega$, permits the separation of the different time scales associated 
with each energy in the HFE. The effective Hamiltonians in the two different regimes 
are:
\begin{multline}
 \Heff^{U\gg\omega}=\mathcal{J}_0(x)\Jpeff\sum\limits_{i=1}^M d^\dagger_{2i}d_{2i-1}
 \\+\mathcal{J}_0(y)\Jpeff\sum\limits_{i=1}^{M-1} d^\dagger_{2i+1}d_{2i}+H.c.
 +\sum\limits_{i=1}^{2M}\mu_in^d_i \ .\\
 \mu_i =  \left\{ \begin{array}{lcc}
		 \Jpeff+\Jeff+U & \text{if} & 1<i<2M\\
		 \Jpeff+U & \text{if} & i\in \{1,2M\}
		   \end{array} \right.\qquad
 \label{model_2}
\end{multline}

\begin{multline}
 \Heff^{\omega\gg U}=\mathcal{J}^2_0(x/2)\Jpeff\sum\limits_{i=1}^M d^\dagger_{2i}d_{2i-1}
 \\+\mathcal{J}^2_0(y/2)\Jpeff\sum\limits_{i=1}^{M-1} d^\dagger_{2i+1}d_{2i}+H.c.
 +\sum\limits_{i=1}^{2M}\mu_in^d_i \ .\\
 \mu_i =  \left\{ \begin{array}{lcc}
		 \mathcal{J}^2_0(x/2)\Jpeff+\mathcal{J}^2_0(y/2)\Jeff+U & \text{if} & 1<i<2M\\
		 \mathcal{J}^2_0(x/2)\Jpeff+U & \text{if} & i\in \{1,2M\}
		   \end{array} \right.
\end{multline}
We emphasize that only in the regime $U\gg\omega$ the effect of 
the ac field is to renormalize the hopping parameters but not the 
effective chemical potential. This is a nontrivial point, key to the 
understanding of the Shockley transfer phenomenon.

\section{Acknowledgements} 
We acknowledge F. Hofmann for enlightening discussions. This work was supported by the Spanish
Ministry through Grants No. MAT2014-58241-P. and No. FIS2013-41716-P.

\end{document}

%% file: energyplot.tex
\begingroup
\newcommand{\ft}[0]{\footnotesize}
  \makeatletter
  \providecommand\color[2][]{%
    \GenericError{(gnuplot) \space\space\space\@spaces}{%
      Package color not loaded in conjunction with
      terminal option `colourtext'%
    }{See the gnuplot documentation for explanation.%
    }{Either use 'blacktext' in gnuplot or load the package
      color.sty in LaTeX.}%
    \renewcommand\color[2][]{}%
  }%
  \providecommand\includegraphics[2][]{%
    \GenericError{(gnuplot) \space\space\space\@spaces}{%
      Package graphicx or graphics not loaded%
    }{See the gnuplot documentation for explanation.%
    }{The gnuplot epslatex terminal needs graphicx.sty or graphics.sty.}%
    \renewcommand\includegraphics[2][]{}%
  }%
  \providecommand\rotatebox[2]{#2}%
  \@ifundefined{ifGPcolor}{%
    \newif\ifGPcolor
    \GPcolortrue
  }{}%
  \@ifundefined{ifGPblacktext}{%
    \newif\ifGPblacktext
    \GPblacktextfalse
  }{}%
  \let\gplgaddtomacro\g@addto@macro
  \gdef\gplbacktext{}%
  \gdef\gplfronttext{}%
  \makeatother
  \ifGPblacktext
    \def\colorrgb#1{}%
    \def\colorgray#1{}%
  \else
    \ifGPcolor
      \def\colorrgb#1{\color[rgb]{#1}}%
      \def\colorgray#1{\color[gray]{#1}}%
      \expandafter\def\csname LTw\endcsname{\color{white}}%
      \expandafter\def\csname LTb\endcsname{\color{black}}%
      \expandafter\def\csname LTa\endcsname{\color{black}}%
      \expandafter\def\csname LT0\endcsname{\color[rgb]{1,0,0}}%
      \expandafter\def\csname LT1\endcsname{\color[rgb]{0,1,0}}%
      \expandafter\def\csname LT2\endcsname{\color[rgb]{0,0,1}}%
      \expandafter\def\csname LT3\endcsname{\color[rgb]{1,0,1}}%
      \expandafter\def\csname LT4\endcsname{\color[rgb]{0,1,1}}%
      \expandafter\def\csname LT5\endcsname{\color[rgb]{1,1,0}}%
      \expandafter\def\csname LT6\endcsname{\color[rgb]{0,0,0}}%
      \expandafter\def\csname LT7\endcsname{\color[rgb]{1,0.3,0}}%
      \expandafter\def\csname LT8\endcsname{\color[rgb]{0.5,0.5,0.5}}%
    \else
      \def\colorrgb#1{\color{black}}%
      \def\colorgray#1{\color[gray]{#1}}%
      \expandafter\def\csname LTw\endcsname{\color{white}}%
      \expandafter\def\csname LTb\endcsname{\color{black}}%
      \expandafter\def\csname LTa\endcsname{\color{black}}%
      \expandafter\def\csname LT0\endcsname{\color{black}}%
      \expandafter\def\csname LT1\endcsname{\color{black}}%
      \expandafter\def\csname LT2\endcsname{\color{black}}%
      \expandafter\def\csname LT3\endcsname{\color{black}}%
      \expandafter\def\csname LT4\endcsname{\color{black}}%
      \expandafter\def\csname LT5\endcsname{\color{black}}%
      \expandafter\def\csname LT6\endcsname{\color{black}}%
      \expandafter\def\csname LT7\endcsname{\color{black}}%
      \expandafter\def\csname LT8\endcsname{\color{black}}%
    \fi
  \fi
  \setlength{\unitlength}{0.0500bp}%
  \begin{picture}(4874.00,3400.00)%
    \gplgaddtomacro\gplbacktext{%
      \csname LTb\endcsname%
      \put(452,2069){\makebox(0,0)[r]{\strut{}\ft0}}%
      \put(452,2321){\makebox(0,0)[r]{\strut{}\ft2}}%
      \put(452,2574){\makebox(0,0)[r]{\strut{}\ft4}}%
      \put(452,2826){\makebox(0,0)[r]{\strut{}\ft6}}%
      \put(452,3079){\makebox(0,0)[r]{\strut{}\ft8}}%
      \put(452,3331){\makebox(0,0)[r]{\strut{}\ft10}}%
      \put(584,1786){\makebox(0,0){\strut{}}}%
      \put(1559,1786){\makebox(0,0){\strut{}}}%
      \put(2533,1786){\makebox(0,0){\strut{}}}%
      \put(78,2668){\rotatebox{-270}{\makebox(0,0){\strut{}\ft $E_n/\Jeff$}}}%
      \put(1558,1720){\makebox(0,0){\strut{}}}%
      \put(779,3199){\makebox(0,0)[l]{\strut{}\ft a)}}%
    }%
    \gplgaddtomacro\gplfronttext{%
    }%
    \gplgaddtomacro\gplbacktext{%
      \csname LTb\endcsname%
      \put(2646,2069){\makebox(0,0)[r]{\strut{}}}%
      \put(2646,2321){\makebox(0,0)[r]{\strut{}}}%
      \put(2646,2574){\makebox(0,0)[r]{\strut{}}}%
      \put(2646,2826){\makebox(0,0)[r]{\strut{}}}%
      \put(2646,3079){\makebox(0,0)[r]{\strut{}}}%
      \put(2646,3331){\makebox(0,0)[r]{\strut{}}}%
      \put(2778,1786){\makebox(0,0){\strut{}}}%
      \put(3752,1786){\makebox(0,0){\strut{}}}%
      \put(4726,1786){\makebox(0,0){\strut{}}}%
      \put(2888,2668){\rotatebox{-270}{\makebox(0,0){\strut{}}}}%
      \put(3752,1720){\makebox(0,0){\strut{}}}%
      \put(2973,3199){\makebox(0,0)[l]{\strut{}\ft b)}}%
    }%
    \gplgaddtomacro\gplfronttext{%
    }%
    \gplgaddtomacro\gplbacktext{%
      \csname LTb\endcsname%
      \put(452,588){\makebox(0,0)[r]{\strut{}\ft0}}%
      \put(452,900){\makebox(0,0)[r]{\strut{}\ft2}}%
      \put(452,1211){\makebox(0,0)[r]{\strut{}\ft4}}%
      \put(452,1523){\makebox(0,0)[r]{\strut{}\ft6}}%
      \put(452,1835){\makebox(0,0)[r]{\strut{}\ft8}}%
      \put(584,290){\makebox(0,0){\strut{}\ft0}}%
      \put(1559,290){\makebox(0,0){\strut{}\ft1}}%
      \put(2533,290){\makebox(0,0){\strut{}\ft2}}%
      \put(78,1172){\rotatebox{-270}{\makebox(0,0){\strut{}\ft $E_n/\Jeff$}}}%
      \put(1558,-40){\makebox(0,0){\strut{}\ft $\lambda$}}%
      \put(779,1703){\makebox(0,0)[l]{\strut{}\ft c)}}%
    }%
    \gplgaddtomacro\gplfronttext{%
    }%
    \gplgaddtomacro\gplbacktext{%
      \csname LTb\endcsname%
      \put(2646,588){\makebox(0,0)[r]{\strut{}}}%
      \put(2646,900){\makebox(0,0)[r]{\strut{}}}%
      \put(2646,1211){\makebox(0,0)[r]{\strut{}}}%
      \put(2646,1523){\makebox(0,0)[r]{\strut{}}}%
      \put(2646,1835){\makebox(0,0)[r]{\strut{}}}%
      \put(2778,290){\makebox(0,0){\strut{}\ft0}}%
      \put(3752,290){\makebox(0,0){\strut{}\ft1}}%
      \put(4726,290){\makebox(0,0){\strut{}\ft2}}%
      \put(2756,1172){\rotatebox{-270}{\makebox(0,0){\strut{}}}}%
      \put(3752,-40){\makebox(0,0){\strut{}\ft $\lambda$}}%
      \put(2973,1703){\makebox(0,0)[l]{\strut{}\ft d)}}%
    }%
    \gplgaddtomacro\gplfronttext{%
    }%
    \gplbacktext
    \put(0,0){\includegraphics{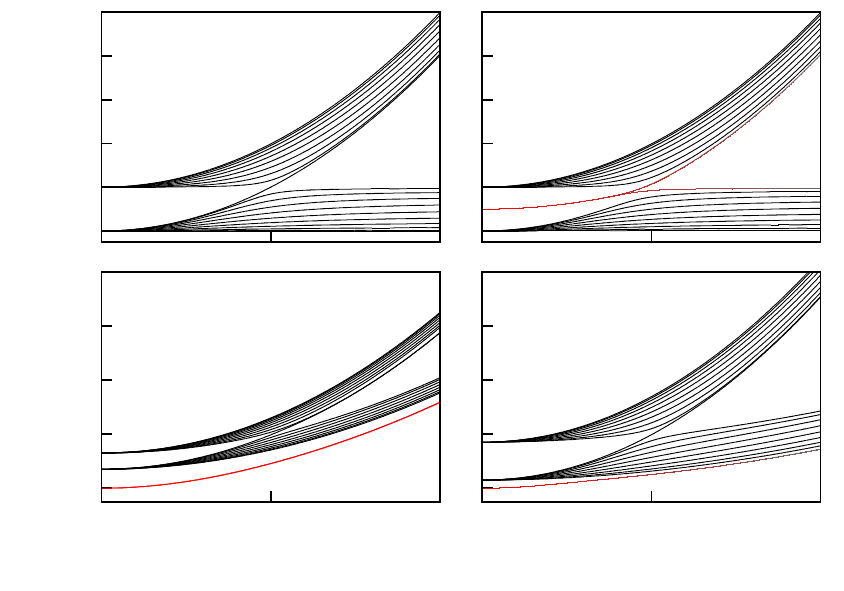}}%
    \gplfronttext
  \end{picture}%
\endgroup

%% file: Dynamics_nuevo.tex
\begingroup
\newcommand{\ft}[0]{\footnotesize}
  \makeatletter
  \providecommand\color[2][]{%
    \GenericError{(gnuplot) \space\space\space\@spaces}{%
      Package color not loaded in conjunction with
      terminal option `colourtext'%
    }{See the gnuplot documentation for explanation.%
    }{Either use 'blacktext' in gnuplot or load the package
      color.sty in LaTeX.}%
    \renewcommand\color[2][]{}%
  }%
  \providecommand\includegraphics[2][]{%
    \GenericError{(gnuplot) \space\space\space\@spaces}{%
      Package graphicx or graphics not loaded%
    }{See the gnuplot documentation for explanation.%
    }{The gnuplot epslatex terminal needs graphicx.sty or graphics.sty.}%
    \renewcommand\includegraphics[2][]{}%
  }%
  \providecommand\rotatebox[2]{#2}%
  \@ifundefined{ifGPcolor}{%
    \newif\ifGPcolor
    \GPcolortrue
  }{}%
  \@ifundefined{ifGPblacktext}{%
    \newif\ifGPblacktext
    \GPblacktextfalse
  }{}%
  \let\gplgaddtomacro\g@addto@macro
  \gdef\gplbacktext{}%
  \gdef\gplfronttext{}%
  \makeatother
  \ifGPblacktext
    \def\colorrgb#1{}%
    \def\colorgray#1{}%
  \else
    \ifGPcolor
      \def\colorrgb#1{\color[rgb]{#1}}%
      \def\colorgray#1{\color[gray]{#1}}%
      \expandafter\def\csname LTw\endcsname{\color{white}}%
      \expandafter\def\csname LTb\endcsname{\color{black}}%
      \expandafter\def\csname LTa\endcsname{\color{black}}%
      \expandafter\def\csname LT0\endcsname{\color[rgb]{1,0,0}}%
      \expandafter\def\csname LT1\endcsname{\color[rgb]{0,1,0}}%
      \expandafter\def\csname LT2\endcsname{\color[rgb]{0,0,1}}%
      \expandafter\def\csname LT3\endcsname{\color[rgb]{1,0,1}}%
      \expandafter\def\csname LT4\endcsname{\color[rgb]{0,1,1}}%
      \expandafter\def\csname LT5\endcsname{\color[rgb]{1,1,0}}%
      \expandafter\def\csname LT6\endcsname{\color[rgb]{0,0,0}}%
      \expandafter\def\csname LT7\endcsname{\color[rgb]{1,0.3,0}}%
      \expandafter\def\csname LT8\endcsname{\color[rgb]{0.5,0.5,0.5}}%
    \else
      \def\colorrgb#1{\color{black}}%
      \def\colorgray#1{\color[gray]{#1}}%
      \expandafter\def\csname LTw\endcsname{\color{white}}%
      \expandafter\def\csname LTb\endcsname{\color{black}}%
      \expandafter\def\csname LTa\endcsname{\color{black}}%
      \expandafter\def\csname LT0\endcsname{\color{black}}%
      \expandafter\def\csname LT1\endcsname{\color{black}}%
      \expandafter\def\csname LT2\endcsname{\color{black}}%
      \expandafter\def\csname LT3\endcsname{\color{black}}%
      \expandafter\def\csname LT4\endcsname{\color{black}}%
      \expandafter\def\csname LT5\endcsname{\color{black}}%
      \expandafter\def\csname LT6\endcsname{\color{black}}%
      \expandafter\def\csname LT7\endcsname{\color{black}}%
      \expandafter\def\csname LT8\endcsname{\color{black}}%
    \fi
  \fi
  \setlength{\unitlength}{0.0500bp}%
  \begin{picture}(4874.00,6802.00)%
    \gplgaddtomacro\gplbacktext{%
      \csname LTb\endcsname%
      \put(599,4965){\makebox(0,0)[r]{\strut{}\ft 0}}%
      \put(599,6426){\makebox(0,0)[r]{\strut{}\ft $5\cdot 10^4$}}%
      \put(950,4811){\makebox(0,0){\strut{}\ft2}}%
      \put(1242,4811){\makebox(0,0){\strut{}\ft4}}%
      \put(1535,4811){\makebox(0,0){\strut{}\ft6}}%
      \put(1827,4811){\makebox(0,0){\strut{}\ft8}}%
      \put(2119,4811){\makebox(0,0){\strut{}\ft10}}%
      \put(-39,5695){\rotatebox{-270}{\makebox(0,0){\strut{}\ft Time $[J^{-1}]$}}}%
      \put(1461,4459){\makebox(0,0){\strut{}}}%
    }%
    \gplgaddtomacro\gplfronttext{%
      \csname LTb\endcsname%
      \put(746,6338){\makebox(0,0)[l]{\strut{}\ft a)}}%
    }%
    \gplgaddtomacro\gplbacktext{%
      \csname LTb\endcsname%
      \put(2792,4965){\makebox(0,0)[r]{\strut{}\ft 0}}%
      \put(2792,6426){\makebox(0,0)[r]{\strut{}\ft $5\cdot 10^2$}}%
      \put(3143,4811){\makebox(0,0){\strut{}\ft2}}%
      \put(3435,4811){\makebox(0,0){\strut{}\ft4}}%
      \put(3728,4811){\makebox(0,0){\strut{}\ft6}}%
      \put(4020,4811){\makebox(0,0){\strut{}\ft8}}%
      \put(4312,4811){\makebox(0,0){\strut{}\ft10}}%
      \put(2374,5695){\rotatebox{-270}{\makebox(0,0){\strut{}}}}%
      \put(3654,4459){\makebox(0,0){\strut{}}}%
    }%
    \gplgaddtomacro\gplfronttext{%
      \csname LTb\endcsname%
      \put(2939,6338){\makebox(0,0)[l]{\strut{}\ft b)}}%
    }%
    \gplgaddtomacro\gplbacktext{%
      \csname LTb\endcsname%
      \put(599,3162){\makebox(0,0)[r]{\strut{}\ft 0}}%
      \put(599,4623){\makebox(0,0)[r]{\strut{}\ft $10^5$}}%
      \put(950,3008){\makebox(0,0){\strut{}\ft2}}%
      \put(1242,3008){\makebox(0,0){\strut{}\ft4}}%
      \put(1535,3008){\makebox(0,0){\strut{}\ft6}}%
      \put(1827,3008){\makebox(0,0){\strut{}\ft8}}%
      \put(2119,3008){\makebox(0,0){\strut{}\ft10}}%
      \put(-39,3892){\rotatebox{-270}{\makebox(0,0){\strut{}\ft Time $[J^{-1}]$}}}%
      \put(1461,2656){\makebox(0,0){\strut{}}}%
    }%
    \gplgaddtomacro\gplfronttext{%
      \csname LTb\endcsname%
      \put(746,4535){\makebox(0,0)[l]{\strut{}\ft c)}}%
    }%
    \gplgaddtomacro\gplbacktext{%
      \csname LTb\endcsname%
      \put(2792,3162){\makebox(0,0)[r]{\strut{}\ft 0}}%
      \put(2792,4623){\makebox(0,0)[r]{\strut{}\ft $10^5$}}%
      \put(2990,3008){\makebox(0,0){\strut{}\ft1}}%
      \put(3256,3008){\makebox(0,0){\strut{}\ft3}}%
      \put(3522,3008){\makebox(0,0){\strut{}\ft5}}%
      \put(3787,3008){\makebox(0,0){\strut{}\ft7}}%
      \put(4053,3008){\makebox(0,0){\strut{}\ft9}}%
      \put(4319,3008){\makebox(0,0){\strut{}\ft11}}%
      \put(2374,3892){\rotatebox{-270}{\makebox(0,0){\strut{}}}}%
      \put(3654,2656){\makebox(0,0){\strut{}}}%
    }%
    \gplgaddtomacro\gplfronttext{%
      \csname LTb\endcsname%
      \put(2939,4535){\makebox(0,0)[l]{\strut{}\ft d)}}%
    }%
    \gplgaddtomacro\gplbacktext{%
      \csname LTb\endcsname%
      \put(599,1360){\makebox(0,0)[r]{\strut{}\ft 0}}%
      \put(599,2821){\makebox(0,0)[r]{\strut{}\ft $2\cdot 10^4$}}%
      \put(950,1206){\makebox(0,0){\strut{}\ft2}}%
      \put(1242,1206){\makebox(0,0){\strut{}\ft4}}%
      \put(1535,1206){\makebox(0,0){\strut{}\ft6}}%
      \put(1827,1206){\makebox(0,0){\strut{}\ft8}}%
      \put(2119,1206){\makebox(0,0){\strut{}\ft10}}%
      \put(-39,2090){\rotatebox{-270}{\makebox(0,0){\strut{}\ft Time $[J^{-1}]$}}}%
      \put(1461,876){\makebox(0,0){\strut{}\ft Lattice site}}%
    }%
    \gplgaddtomacro\gplfronttext{%
      \csname LTb\endcsname%
      \put(746,2733){\makebox(0,0)[l]{\strut{}\ft e)}}%
    }%
    \gplgaddtomacro\gplbacktext{%
      \csname LTb\endcsname%
      \put(2792,1360){\makebox(0,0)[r]{\strut{}\ft 0}}%
      \put(2792,2821){\makebox(0,0)[r]{\strut{}\ft $10^3$}}%
      \put(3143,1206){\makebox(0,0){\strut{}\ft2}}%
      \put(3435,1206){\makebox(0,0){\strut{}\ft4}}%
      \put(3728,1206){\makebox(0,0){\strut{}\ft6}}%
      \put(4020,1206){\makebox(0,0){\strut{}\ft8}}%
      \put(4312,1206){\makebox(0,0){\strut{}\ft10}}%
      \put(2770,2090){\rotatebox{-270}{\makebox(0,0){\strut{}}}}%
      \put(3654,876){\makebox(0,0){\strut{}\ft Lattice site}}%
    }%
    \gplgaddtomacro\gplfronttext{%
      \csname LTb\endcsname%
      \put(1462,366){\makebox(0,0){\strut{}\ft0}}%
      \put(1949,366){\makebox(0,0){\strut{}\ft0.5}}%
      \put(2436,366){\makebox(0,0){\strut{}\ft1}}%
      \put(2923,366){\makebox(0,0){\strut{}\ft1.5}}%
      \put(3411,366){\makebox(0,0){\strut{}\ft2}}%
      \put(2436,102){\makebox(0,0){\strut{}\ft Occupation}}%
      \csname LTb\endcsname%
      \put(2939,2733){\makebox(0,0)[l]{\strut{}\ft f)}}%
    }%
    \gplbacktext
    \put(0,0){\includegraphics{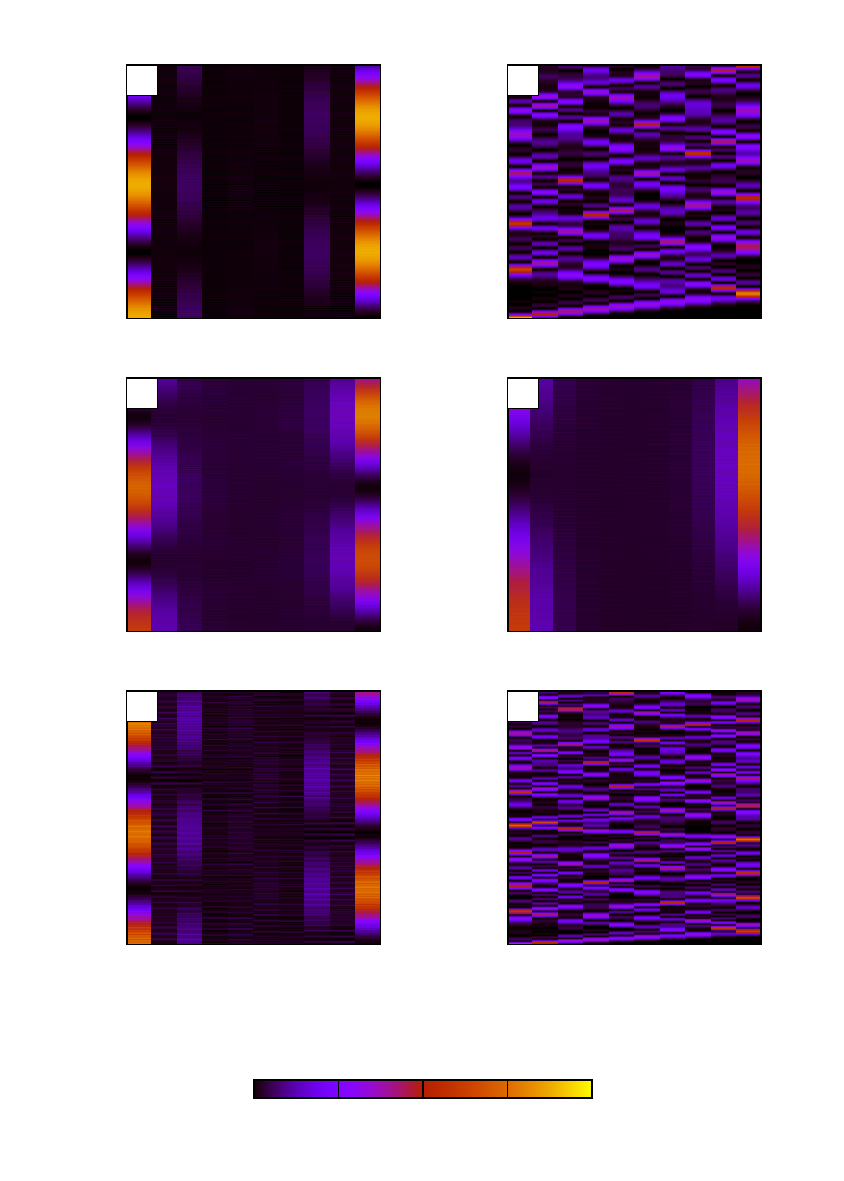}}%
    \gplfronttext
  \end{picture}%
\endgroup

%% file: PhaseE_Ttime.tex
\begingroup
\newcommand{\ft}[0]{\footnotesize}
  \makeatletter
  \providecommand\color[2][]{%
    \GenericError{(gnuplot) \space\space\space\@spaces}{%
      Package color not loaded in conjunction with
      terminal option `colourtext'%
    }{See the gnuplot documentation for explanation.%
    }{Either use 'blacktext' in gnuplot or load the package
      color.sty in LaTeX.}%
    \renewcommand\color[2][]{}%
  }%
  \providecommand\includegraphics[2][]{%
    \GenericError{(gnuplot) \space\space\space\@spaces}{%
      Package graphicx or graphics not loaded%
    }{See the gnuplot documentation for explanation.%
    }{The gnuplot epslatex terminal needs graphicx.sty or graphics.sty.}%
    \renewcommand\includegraphics[2][]{}%
  }%
  \providecommand\rotatebox[2]{#2}%
  \@ifundefined{ifGPcolor}{%
    \newif\ifGPcolor
    \GPcolortrue
  }{}%
  \@ifundefined{ifGPblacktext}{%
    \newif\ifGPblacktext
    \GPblacktextfalse
  }{}%
  \let\gplgaddtomacro\g@addto@macro
  \gdef\gplbacktext{}%
  \gdef\gplfronttext{}%
  \makeatother
  \ifGPblacktext
    \def\colorrgb#1{}%
    \def\colorgray#1{}%
  \else
    \ifGPcolor
      \def\colorrgb#1{\color[rgb]{#1}}%
      \def\colorgray#1{\color[gray]{#1}}%
      \expandafter\def\csname LTw\endcsname{\color{white}}%
      \expandafter\def\csname LTb\endcsname{\color{black}}%
      \expandafter\def\csname LTa\endcsname{\color{black}}%
      \expandafter\def\csname LT0\endcsname{\color[rgb]{1,0,0}}%
      \expandafter\def\csname LT1\endcsname{\color[rgb]{0,1,0}}%
      \expandafter\def\csname LT2\endcsname{\color[rgb]{0,0,1}}%
      \expandafter\def\csname LT3\endcsname{\color[rgb]{1,0,1}}%
      \expandafter\def\csname LT4\endcsname{\color[rgb]{0,1,1}}%
      \expandafter\def\csname LT5\endcsname{\color[rgb]{1,1,0}}%
      \expandafter\def\csname LT6\endcsname{\color[rgb]{0,0,0}}%
      \expandafter\def\csname LT7\endcsname{\color[rgb]{1,0.3,0}}%
      \expandafter\def\csname LT8\endcsname{\color[rgb]{0.5,0.5,0.5}}%
    \else
      \def\colorrgb#1{\color{black}}%
      \def\colorgray#1{\color[gray]{#1}}%
      \expandafter\def\csname LTw\endcsname{\color{white}}%
      \expandafter\def\csname LTb\endcsname{\color{black}}%
      \expandafter\def\csname LTa\endcsname{\color{black}}%
      \expandafter\def\csname LT0\endcsname{\color{black}}%
      \expandafter\def\csname LT1\endcsname{\color{black}}%
      \expandafter\def\csname LT2\endcsname{\color{black}}%
      \expandafter\def\csname LT3\endcsname{\color{black}}%
      \expandafter\def\csname LT4\endcsname{\color{black}}%
      \expandafter\def\csname LT5\endcsname{\color{black}}%
      \expandafter\def\csname LT6\endcsname{\color{black}}%
      \expandafter\def\csname LT7\endcsname{\color{black}}%
      \expandafter\def\csname LT8\endcsname{\color{black}}%
    \fi
  \fi
  \setlength{\unitlength}{0.0500bp}%
  \begin{picture}(4874.00,2550.00)%
    \gplgaddtomacro\gplbacktext{%
      \csname LTb\endcsname%
      \put(842,1020){\makebox(0,0)[r]{\strut{}\ft 0.1}}%
      \put(842,1670){\makebox(0,0)[r]{\strut{}\ft 0.5}}%
      \put(842,2482){\makebox(0,0)[r]{\strut{}\ft 1}}%
      \put(974,844){\makebox(0,0){\strut{}\ft 0.2}}%
      \put(1624,844){\makebox(0,0){\strut{}\ft 1}}%
      \put(2436,844){\makebox(0,0){\strut{}\ft 2}}%
      \put(270,1751){\rotatebox{-270}{\makebox(0,0){\strut{}\ft $|\mathcal{J}_0(Ea_0/\omega)|$}}}%
      \put(1705,624){\makebox(0,0){\strut{}\ft $\lambda$}}%
    }%
    \gplgaddtomacro\gplfronttext{%
      \csname LTb\endcsname%
      \put(974,267){\makebox(0,0){\strut{}\ft0}}%
      \put(2192,267){\makebox(0,0){\strut{}\ft0.5}}%
      \put(1583,0){\makebox(0,0){\strut{}\ft $\mathcal{G}\left[\lvert\psi\rangle\right]$}}%
      \put(2233,2320){\makebox(0,0)[l]{\strut{}\ft \textcolor{white}{a)}}}%
    }%
    \gplgaddtomacro\gplbacktext{%
      \csname LTb\endcsname%
      \put(842,1020){\makebox(0,0)[r]{\strut{}\ft 0.1}}%
      \put(842,1670){\makebox(0,0)[r]{\strut{}\ft 0.5}}%
      \put(842,2482){\makebox(0,0)[r]{\strut{}\ft 1}}%
      \put(974,844){\makebox(0,0){\strut{}\ft 0.2}}%
      \put(1624,844){\makebox(0,0){\strut{}\ft 1}}%
      \put(2436,844){\makebox(0,0){\strut{}\ft 2}}%
      \put(270,1751){\rotatebox{-270}{\makebox(0,0){\strut{}\ft $|\mathcal{J}_0(Ea_0/\omega)|$}}}%
      \put(1705,624){\makebox(0,0){\strut{}\ft $\lambda$}}%
    }%
    \gplgaddtomacro\gplfronttext{%
      \csname LTb\endcsname%
      \put(2233,2320){\makebox(0,0)[l]{\strut{}\ft \textcolor{white}{a)}}}%
    }%
    \gplgaddtomacro\gplbacktext{%
      \csname LTb\endcsname%
      \put(2792,1020){\makebox(0,0)[r]{\strut{}\ft 0.1}}%
      \put(2792,1669){\makebox(0,0)[r]{\strut{}\ft 0.5}}%
      \put(2792,2481){\makebox(0,0)[r]{\strut{}\ft 1}}%
      \put(2924,844){\makebox(0,0){\strut{}\ft 0.2}}%
      \put(3573,844){\makebox(0,0){\strut{}\ft 1}}%
      \put(4385,844){\makebox(0,0){\strut{}\ft 2}}%
      \put(2440,1750){\rotatebox{-270}{\makebox(0,0){\strut{}}}}%
      \put(3654,624){\makebox(0,0){\strut{}\ft $\lambda$}}%
    }%
    \gplgaddtomacro\gplfronttext{%
      \csname LTb\endcsname%
      \put(2807,267){\makebox(0,0){\strut{}$0$}}%
      \put(3233,267){\makebox(0,0){\strut{}$4$}}%
      \put(3659,267){\makebox(0,0){\strut{}$8$}}%
      \put(4085,267){\makebox(0,0){\strut{}$12$}}%
      \put(4512,267){\makebox(0,0){\strut{}\ft $>16$}}%
      \put(3659,0){\makebox(0,0){\strut{}\ft $\log_{10}(T_0 J)$}}%
      \put(4182,2319){\makebox(0,0)[l]{\strut{}\ft \textcolor{black}{b)}}}%
    }%
    \gplgaddtomacro\gplbacktext{%
      \csname LTb\endcsname%
      \put(2792,1020){\makebox(0,0)[r]{\strut{}\ft 0.1}}%
      \put(2792,1669){\makebox(0,0)[r]{\strut{}\ft 0.5}}%
      \put(2792,2481){\makebox(0,0)[r]{\strut{}\ft 1}}%
      \put(2924,844){\makebox(0,0){\strut{}\ft 0.2}}%
      \put(3573,844){\makebox(0,0){\strut{}\ft 1}}%
      \put(4385,844){\makebox(0,0){\strut{}\ft 2}}%
      \put(2440,1750){\rotatebox{-270}{\makebox(0,0){\strut{}}}}%
      \put(3654,624){\makebox(0,0){\strut{}\ft $\lambda$}}%
    }%
    \gplgaddtomacro\gplfronttext{%
      \csname LTb\endcsname%
      \put(4182,2319){\makebox(0,0)[l]{\strut{}\ft \textcolor{black}{b)}}}%
    }%
    \gplbacktext
    \put(0,0){\includegraphics{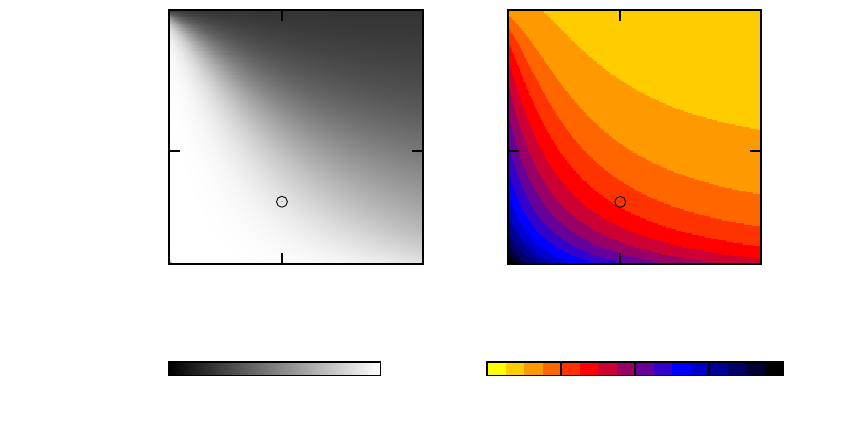}}%
    \gplfronttext
  \end{picture}%
\endgroup

%% file: quasiener_paper_nuevo.tex
\begingroup
\newcommand{\ft}[0]{\footnotesize}
  \makeatletter
  \providecommand\color[2][]{%
    \GenericError{(gnuplot) \space\space\space\@spaces}{%
      Package color not loaded in conjunction with
      terminal option `colourtext'%
    }{See the gnuplot documentation for explanation.%
    }{Either use 'blacktext' in gnuplot or load the package
      color.sty in LaTeX.}%
    \renewcommand\color[2][]{}%
  }%
  \providecommand\includegraphics[2][]{%
    \GenericError{(gnuplot) \space\space\space\@spaces}{%
      Package graphicx or graphics not loaded%
    }{See the gnuplot documentation for explanation.%
    }{The gnuplot epslatex terminal needs graphicx.sty or graphics.sty.}%
    \renewcommand\includegraphics[2][]{}%
  }%
  \providecommand\rotatebox[2]{#2}%
  \@ifundefined{ifGPcolor}{%
    \newif\ifGPcolor
    \GPcolortrue
  }{}%
  \@ifundefined{ifGPblacktext}{%
    \newif\ifGPblacktext
    \GPblacktextfalse
  }{}%
  \let\gplgaddtomacro\g@addto@macro
  \gdef\gplbacktext{}%
  \gdef\gplfronttext{}%
  \makeatother
  \ifGPblacktext
    \def\colorrgb#1{}%
    \def\colorgray#1{}%
  \else
    \ifGPcolor
      \def\colorrgb#1{\color[rgb]{#1}}%
      \def\colorgray#1{\color[gray]{#1}}%
      \expandafter\def\csname LTw\endcsname{\color{white}}%
      \expandafter\def\csname LTb\endcsname{\color{black}}%
      \expandafter\def\csname LTa\endcsname{\color{black}}%
      \expandafter\def\csname LT0\endcsname{\color[rgb]{1,0,0}}%
      \expandafter\def\csname LT1\endcsname{\color[rgb]{0,1,0}}%
      \expandafter\def\csname LT2\endcsname{\color[rgb]{0,0,1}}%
      \expandafter\def\csname LT3\endcsname{\color[rgb]{1,0,1}}%
      \expandafter\def\csname LT4\endcsname{\color[rgb]{0,1,1}}%
      \expandafter\def\csname LT5\endcsname{\color[rgb]{1,1,0}}%
      \expandafter\def\csname LT6\endcsname{\color[rgb]{0,0,0}}%
      \expandafter\def\csname LT7\endcsname{\color[rgb]{1,0.3,0}}%
      \expandafter\def\csname LT8\endcsname{\color[rgb]{0.5,0.5,0.5}}%
    \else
      \def\colorrgb#1{\color{black}}%
      \def\colorgray#1{\color[gray]{#1}}%
      \expandafter\def\csname LTw\endcsname{\color{white}}%
      \expandafter\def\csname LTb\endcsname{\color{black}}%
      \expandafter\def\csname LTa\endcsname{\color{black}}%
      \expandafter\def\csname LT0\endcsname{\color{black}}%
      \expandafter\def\csname LT1\endcsname{\color{black}}%
      \expandafter\def\csname LT2\endcsname{\color{black}}%
      \expandafter\def\csname LT3\endcsname{\color{black}}%
      \expandafter\def\csname LT4\endcsname{\color{black}}%
      \expandafter\def\csname LT5\endcsname{\color{black}}%
      \expandafter\def\csname LT6\endcsname{\color{black}}%
      \expandafter\def\csname LT7\endcsname{\color{black}}%
      \expandafter\def\csname LT8\endcsname{\color{black}}%
    \fi
  \fi
  \setlength{\unitlength}{0.0500bp}%
  \begin{picture}(4874.00,4534.00)%
    \gplgaddtomacro\gplbacktext{%
      \csname LTb\endcsname%
      \put(672,2947){\makebox(0,0)[r]{\strut{}\ft 0}}%
      \put(672,3312){\makebox(0,0)[r]{\strut{}\ft 5}}%
      \put(672,3678){\makebox(0,0)[r]{\strut{}\ft 10}}%
      \put(672,4043){\makebox(0,0)[r]{\strut{}\ft 15}}%
      \put(672,4408){\makebox(0,0)[r]{\strut{}\ft 20}}%
      \put(804,2749){\makebox(0,0){\strut{}\ft 0}}%
      \put(1535,2749){\makebox(0,0){\strut{}\ft 0.5}}%
      \put(2265,2749){\makebox(0,0){\strut{}\ft 1}}%
      \put(166,3677){\rotatebox{-270}{\makebox(0,0){\strut{}\ft $2Ea_0/\omega$}}}%
      \put(1534,2463){\makebox(0,0){\strut{}\ft $b_0$ $(a_0)$}}%
    }%
    \gplgaddtomacro\gplfronttext{%
      \csname LTb\endcsname%
      \put(811,4317){\makebox(0,0)[l]{\strut{}\ft a)}}%
    }%
    \gplgaddtomacro\gplbacktext{%
      \csname LTb\endcsname%
      \put(2378,2947){\makebox(0,0)[r]{\strut{}}}%
      \put(2378,3312){\makebox(0,0)[r]{\strut{}}}%
      \put(2378,3678){\makebox(0,0)[r]{\strut{}}}%
      \put(2378,4043){\makebox(0,0)[r]{\strut{}}}%
      \put(2378,4408){\makebox(0,0)[r]{\strut{}}}%
      \put(2510,2749){\makebox(0,0){\strut{}\ft 0}}%
      \put(3241,2749){\makebox(0,0){\strut{}\ft 0.5}}%
      \put(3971,2749){\makebox(0,0){\strut{}\ft 1}}%
      \put(2620,3677){\rotatebox{-270}{\makebox(0,0){\strut{}}}}%
      \put(3240,2463){\makebox(0,0){\strut{}\ft $b_0$ $(a_0)$}}%
    }%
    \gplgaddtomacro\gplfronttext{%
      \csname LTb\endcsname%
      \put(4371,2946){\makebox(0,0)[l]{\strut{}\ft 0}}%
      \put(4371,4405){\makebox(0,0)[l]{\strut{}\ft 0.5}}%
      \put(4702,3675){\rotatebox{-270}{\makebox(0,0){\strut{}\ft $\mathcal{G}\left[\lvert\psi\rangle\right]$}}}%
      \csname LTb\endcsname%
      \put(2517,4317){\makebox(0,0)[l]{\strut{}\ft b)}}%
    }%
    \gplgaddtomacro\gplbacktext{%
      \csname LTb\endcsname%
      \put(2378,2947){\makebox(0,0)[r]{\strut{}}}%
      \put(2378,3312){\makebox(0,0)[r]{\strut{}}}%
      \put(2378,3678){\makebox(0,0)[r]{\strut{}}}%
      \put(2378,4043){\makebox(0,0)[r]{\strut{}}}%
      \put(2378,4408){\makebox(0,0)[r]{\strut{}}}%
      \put(2510,2749){\makebox(0,0){\strut{}\ft 0}}%
      \put(3241,2749){\makebox(0,0){\strut{}\ft 0.5}}%
      \put(3971,2749){\makebox(0,0){\strut{}\ft 1}}%
      \put(2620,3677){\rotatebox{-270}{\makebox(0,0){\strut{}}}}%
      \put(3240,2463){\makebox(0,0){\strut{}\ft $b_0$ $(a_0)$}}%
    }%
    \gplgaddtomacro\gplfronttext{%
      \csname LTb\endcsname%
      \put(2517,4317){\makebox(0,0)[l]{\strut{}\ft b)}}%
    }%
    \gplgaddtomacro\gplbacktext{%
      \csname LTb\endcsname%
      \put(2378,2947){\makebox(0,0)[r]{\strut{}}}%
      \put(2378,3312){\makebox(0,0)[r]{\strut{}}}%
      \put(2378,3678){\makebox(0,0)[r]{\strut{}}}%
      \put(2378,4043){\makebox(0,0)[r]{\strut{}}}%
      \put(2378,4408){\makebox(0,0)[r]{\strut{}}}%
      \put(2510,2749){\makebox(0,0){\strut{}\ft 0}}%
      \put(3241,2749){\makebox(0,0){\strut{}\ft 0.5}}%
      \put(3971,2749){\makebox(0,0){\strut{}\ft 1}}%
      \put(2620,3677){\rotatebox{-270}{\makebox(0,0){\strut{}}}}%
      \put(3240,2463){\makebox(0,0){\strut{}\ft $b_0$ $(a_0)$}}%
    }%
    \gplgaddtomacro\gplfronttext{%
      \csname LTb\endcsname%
      \put(2517,4317){\makebox(0,0)[l]{\strut{}\ft b)}}%
    }%
    \gplgaddtomacro\gplbacktext{%
      \csname LTb\endcsname%
      \put(672,598){\makebox(0,0)[r]{\strut{}}}%
      \put(672,867){\makebox(0,0)[r]{\strut{}}}%
      \put(672,1136){\makebox(0,0)[r]{\strut{}}}%
      \put(672,1405){\makebox(0,0)[r]{\strut{}}}%
      \put(672,1674){\makebox(0,0)[r]{\strut{}}}%
      \put(672,1943){\makebox(0,0)[r]{\strut{}}}%
      \put(672,2212){\makebox(0,0)[r]{\strut{}}}%
      \put(804,346){\makebox(0,0){\strut{}\ft 0}}%
      \put(1699,346){\makebox(0,0){\strut{}\ft 5}}%
      \put(2595,346){\makebox(0,0){\strut{}\ft 10}}%
      \put(3490,346){\makebox(0,0){\strut{}\ft 15}}%
      \put(4385,346){\makebox(0,0){\strut{}\ft 20}}%
      \put(430,1405){\rotatebox{-270}{\makebox(0,0){\strut{}\ft $\epsilon_n$ $[J]$}}}%
      \put(2594,60){\makebox(0,0){\strut{}\ft $2Ea_0/\omega$}}%
    }%
    \gplgaddtomacro\gplfronttext{%
      \csname LTb\endcsname%
      \put(4116,2078){\makebox(0,0)[l]{\strut{}\ft c)}}%
    }%
    \gplbacktext
    \put(0,0){\includegraphics{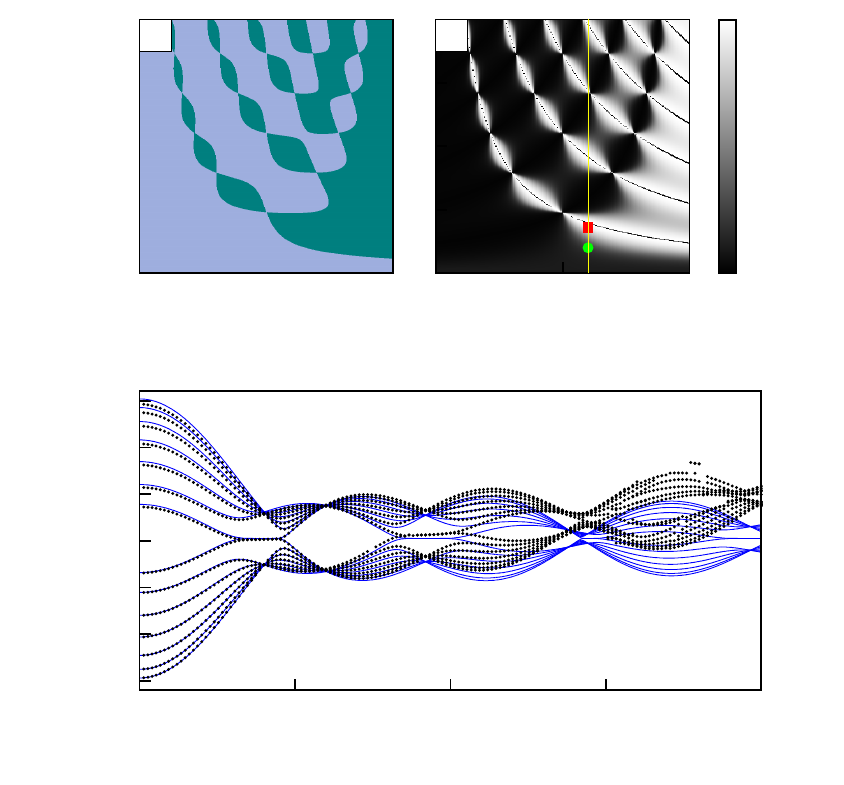}}%
    \gplfronttext
  \end{picture}%
\endgroup